\documentclass[preprint]{JHEP3} 

\usepackage{epsfig,multicol,bbm}

\newcommand\fverb{\setbox\pippobox=\hbox\bgroup\verb}
\newcommand\fverbdo{\egroup\medskip\noindent%
			\fbox{\unhbox\pippobox}\ }
\newcommand\fverbit{\egroup\item[\fbox{\unhbox\pippobox}]}
\newbox\pippobox

\title{Chiral-Yang-Mills theory, non commutative differential geometry,
and the need for a Lie super-algebra}

\author{Jean Thierry-Mieg\\
NCBI. NLM. NIH Bldg 38A\\
8600 Rockville Pike\\
Bethesda, MD 20894, USA\\
Tel   1 (301) 435 49 21   Fax   1 (301) 480 92 41
	E-mail: \email{mieg@ncbi.nlm.nih.gov}\\
and\\ 
Laboratoire de Physique Th\'eorique et d'Astro-particules,\\
CNRS, Montpellier, France.}
\abstract{In Yang-Mills theory, the charges of the left and right 
massless Fermions are independent of each other. We propose a new 
paradigm where 
we remove this freedom and densify the algebraic structure
of Yang-Mills theory by integrating the scalar Higgs field into
a new gauge-chiral 1-form which connects Fermions 
of opposite chiralities. Using the Bianchi identity, we
prove that the corresponding covariant
differential is associative if and only if we gauge
a Lie-Kac super-algebra. In this model, spontaneous symmetry breakdown 
naturally occurs along an odd generator of the super-algebra and 
induces a representation of the Connes-Lott non commutative differential 
geometry of the 2-point finite space.
}

\keywords{SU(2/1), super-algebra , chiral-Yang Mills theory
, Bianchi identity , non commutative differential geometry 
, chirality , standard model
}

 \preprint{2006-01/01}
\dedicated{Dedicated to Yuval Ne'eman on the occasion of his $80^{th}$ birthday.}

\newcommand{\BE}{\begin{equation}}
\newcommand{\EE}{\end{equation}}
\newcommand{\BQ}{\begin{equation} \begin{array}{c}}
\newcommand{\EQ}{\end{array}\end{equation}}
\newcommand{\BT}{\begin{theorem}}
\newcommand{\ET}{\end{theorem}}

\newcommand{\bc}{\begin{center}}
\newcommand{\ec}{\end{center}}

\newcommand{\dHH}{\widehat{d}}

\newcommand{\cHH}{\widehat{c}}
\newcommand{\AHH}{\widehat{A}}
\newcommand{\DHH}{\widehat{D}}
\newcommand{\FHH}{\widehat{F}}
\newcommand{\gHH}{\widehat{g}}

\newcommand{\dHS}{\breve{d}}
\newcommand{\AHS}{\breve{A}}
\newcommand{\DHS}{\breve{D}}
\newcommand{\FHS}{\breve{F}}
\newcommand{\PhiHS}{\breve{\Phi}}
\newcommand{\dT}{\widetilde{d}}
\newcommand{\AT}{\widetilde{A}}
\newcommand{\DT}{\widetilde{D}}
\newcommand{\FT}{\widetilde{F}}
\newcommand{\LX}{\Lambda}

\newcommand{\lX}{\lambda}
\newcommand{\lXB}{\overline{\lX}}
\newcommand{\MX}{\lambda}
\newcommand{\MXB}{\overline{\MX}}

\newcommand{\GX}{\Gamma}
\newcommand{\GXB}{\overline{\Gamma}}
\newcommand{\EX}{\epsilon}
\newcommand{\EXB}{\overline{\EX}}
\newcommand{\ALG}{\mathcal{A}}
\newcommand{\LAG}{\mathcal{L}}

\newcommand{\demi}{{1 \over 2}}

\begin{document} 


\section{Introduction}

The standard model of the fundamental interactions contains
three kinds of particles. The vector bosons, which gauge
a local Lie algebra, the massless Fermions, and the Higgs
scalars. The Yang-Mills formalism describes the interactions
of the vectors with all other particles and the theory is
experimentally very accurate. However, there are 2 drawbacks.
The left and right chiral Fermions are independent of each other,
and the status of the Higgs scalars is unclear, except that 
we absolutely need them to break the symmetry and give a mass
to the different particles. In this paper, we propose a new
construction where the Higgs scalar are included in the geometric
connection one-form and prove that the corresponding covariant differential is associative
if and only if this densified field is valued in the adjoint representation of a
Lie-Kac super-algebra and the Fermions are graded by their chirality.

Although in essence the chiral-Yang-Mills theory is a
generalization of the conventional theory, we prefer the
word densification because we increase the number of
algebraic relations between the known particles, and hence
constrain more tightly their quantum numbers, without introducing
any extra field. In this sense, our model is more economical
than super-symmetry, grand unified theories or super-strings.

Our new construction kills two birds with one stone, 
it unexpectedly streamlines the incorporation of the
2-point finite non commutative differential geometry 
of Connes and Lott \cite{CL90,CL91} into Yang-Mills theory
and provides, at long last, a theoretical framework
for the $SU(2/1)$ phenomenological classification
of the elementary particles, pioneered in 1979 by Ne'eman
and Fairlie \cite{N79,F79,NSF05}. 

The paper starts with the definition of the building
blocks of our new construction. In section 2, we recall 
the construction of the Yang-Mills connection. In section 3,
we summarize the geometrical construction
of the BRS differential. In section 4, we recall the Connes-Lott definition
of the non commutative differential geometry of the 2-point
finite space, and then prove that the associated differential is
algebraic. Then, in section 5, we show how to modify the Yang-Mills connection and
include the Higgs field in a densified chiral-Yang-Mills gauge field
connecting the Fermions of opposite chiralities and then prove
that the corresponding covariant differential is associative
if and only if this densified field is valued in the adjoint representation of a
Lie-Kac super-algebra. In section 6, we show that the chiral-Yang-Mills model naturally 
integrates the non commutative differential geometry of the 2-point
finite space, but contradicts the Connes-Lott analytic presentation \cite{CL90,CL91}.
The differences are discussed in section 7: we base our construction 
on the structure of a finite dimensional
classical Lie super-algebra, the $SU(m/n)$, $OSp(m/n)$, $D(2/1;\alpha)$, $F(4)$ and
$G(3)$ series of Kac, whereas Connes and Lott rely on the very
different notion of a K-cycle over an associative algebra, which forces them to
deviate from the usual formalism of Quantum Field Theory and
invoke the Dixmier trace. In section 8, we generalize the Yang-Mills topological
Lagrangian $F \wedge F$ to the chiral case and show that it has the
same symmetry as the super-Killing metric of the underlying super-algebra.

We will discuss the application of this formalism to the
$SU(2/1)$ model in a separate paper.

\section{The usual Yang-Mills connection one form}

Everything in this section is well known, however
this presentation will allow us to highlight the crucial steps of
the standard construction of Yang-Mills theory to use it
later as a reference when we generalize the theory. Consider a
set of $m$ left and $n$ right spinors 
\BE
 \Psi = (\psi_{L}^{\alpha} , \psi_{R}^{\overline{\alpha}})\;,\;\;\;\alpha = 1, 2 ... m\; ;
\overline{\alpha} = 1, 2... n \, .
\EE
Each $\psi$ spinor is a 2-dimensional column vector. The $\alpha$ and $\overline{\alpha}$ indices 
enumerate the (weak, electric, color) charges. We now consider the eight Hermitian Pauli matrices 
$\sigma_{\mu}$ and  $\overline{\sigma}_{\nu}$, 
$\mu , \nu = 0,1,2,3$ which map the right on the left spinors and vice-versa
\BE
\sigma :  \psi_R \rightarrow \psi_L \,,\qquad \overline{\sigma} :  \psi_L  \rightarrow \psi_R \,.
\EE
Notice that we can multiply $\sigma$ by $\overline{\sigma}$, but we
cannot multiply
$\sigma$ by  $\sigma$ or $\overline{\sigma}$ by $\overline{\sigma}$
because the left and right spinor spaces are different.
We can choose a base where the Pauli matrices read:
\BQ
\sigma_0 = \pmatrix{ 1 & 0 \cr 0 & 1}\,,\qquad
\sigma_1 = \pmatrix{ 0 & 1 \cr 1 & 0}\,,\qquad
\sigma_2 = \pmatrix{ 0 & -i \cr i & 0}\,,\qquad
\sigma_3 = \pmatrix{ 1 & 0 \cr 0 & -1}\,,
\\
\overline{\sigma}_0 = \pmatrix{ -1 & 0 \cr 0 & -1}\,,\qquad
\overline{\sigma}_1 = \pmatrix{ 0 & 1 \cr 1 & 0}\,,\qquad
\overline{\sigma}_2 = \pmatrix{ 0 & -i \cr i & 0}\,,\qquad
\overline{\sigma}_3 = \pmatrix{ 1 & 0 \cr 0 & -1}\,,
\EQ
These matrices satisfy the Dirac relation
\BE
 \sigma_{\mu} \overline{\sigma}_{\nu}  + \sigma_{\nu} \overline{\sigma}_{\mu} = 2 g_{\mu \nu} \; \mathbbm{1}\,,\;\;\;\;\; \mu , \nu = 0,1,2,3\,,
\EE
where $g_{\mu \nu}$ is the Minkowski metric $diagonal(-1,1,1,1)$. The $J^L$ and $J^R$
anti-symmetrized products of the $\sigma$ 
\BQ
 J_{\mu \nu}^L = \demi (\sigma_{\mu} \overline{\sigma}_{\nu} - \sigma_{\nu} \overline{\sigma}_{\mu})
\,,\\
 J_{\mu \nu}^R = \demi (\overline{\sigma}_{\mu} \sigma_{\nu} -  \overline{\sigma_{\nu}} \sigma_{\mu}) 
\,,
\EQ
give us two Hermitian conjugated irreducible representations of
the $SO(1,3)$ Lorentz rotation Lie algebra: $J_R = (J_L)^{\dagger}$.
Working in Minkowski rather than in Euclidean space is crucial if we want to
consider an unequal number of left and right spinors. Indeed, in
Minkowski space both the $\sigma$ and the $\overline{\sigma}$ matrices
are Hermitian $\sigma^{\dagger} = \sigma\;\;;\;\;\;
\overline{\sigma}^{\dagger} = \overline{\sigma}$, whereas in Euclidean
space  $\sigma_E$ and $\overline{\sigma}_E$ are Hermitian conjugates of each other
$\sigma_E^{\dagger} = \overline{\sigma}_E$. In physical terms, the
decoupling of the left and the right Fermions and the
fact that the photon has only 2, and not 4, polarization states only
happens for massless particles defined in Minkowski space. At the speed of light, 
the Lorentz
contraction kills time and the direction of propagation. It only
leaves a 2-dimensional transverse space to describe the spin of the
massless particles.

In Yang-Mills theory, it is further assumed that the left and right
spinors fall into independent (possibly reducible) representations
of a Lie algebra $\ALG_0$, given by the block diagonal matrices 
\BE
\Lambda_a =  \pmatrix{  \lX_{a \beta}^{\alpha} & 0 \cr
              0 & \lXB_{a \overline{\beta}}^{\overline{\alpha}}}
\EE
which are closed under commutation
\BE
  [\LX_a , \LX_b ] = \LX_a  \LX_b - \LX_b  \LX_a = f_{a b}^c \; \LX_c
\EE
 and satisfy the Jacobi identity \BE
 [ \LX_a , [\LX_b , \LX_c ]] +  [ \LX_c , [\LX_a , \LX_b ]] + [ \LX_b , [\LX_c , \LX_a ]] = 0
\EE
The $f^a_{bc}$ are complex numbers called the structure constants of the Lie algebra.
Let us now introduce the Cartan exterior differential $d$. Let $x^{\mu}$,  $\mu = 0,1,2,3$
denote a local system of coordinates. We define the symbols $dx^{\mu}$  and assume
that they commute with the coordinates but anti-commute among themselves:
\BE
 x^{\mu} x^{\nu} = x^{\nu} x^{\mu} \,,\qquad x^{\mu} dx^{\nu} = dx^{\nu} x^{\mu}  \,,\qquad dx^{\mu} dx^{\nu} = - dx^{\nu} dx^{\mu}  \,.
\EE
A polynome of degree $p$ in the $dx^{\mu}$ is called a p-form. The Cartan exterior differential 
is defined in terms of the partial derivatives relative
to the coordinates as $d =  dx^{\mu} \; {\partial / \partial x^{\mu}}$. Since the
$dx^{\mu}$ anti-commute and the partial derivatives commute, the Cartan differential
satisfies the fundamental rule
\BE
 d^2 = 0\,.
\EE
The Yang-Mills connection $A$ is defined as a  space-time dependent Cartan differential one-form
valued in the adjoint representation of the Lie algebra
\BE
 A =  d x^{\mu} \;A_{\mu}^a(x) \;\LX_a\,.
\EE
Notice that $A$ is a matrix-valued one-form, but $A^{a}_{\mu}(x)$ is just a set of commuting
complex valued functions. 
Using $A$, we define the covariant exterior differential
\BE
 D = d + A\,.
\EE
$D$ acts separately on the left and right spinors, $d = dx^{\mu} \partial_{\mu}$ acts as a
derivative and $A$ acts by matrix multiplication.
\BE
 D \psi _L ^{\alpha}(x) =  dx^{\mu} \; (\partial _{\mu} \psi_L^{\alpha}(x) + A_{\mu}^{a}(x) \lX_{a \beta}^{\alpha} \psi_L^{\beta}(x) )\,.
\EE We now compute $D D \psi$. The facts that the $dx^{\mu}$
anti-commute and the $\partial_{\mu}$ satisfy the Leibnitz rule imply
that $D D \psi $ does not depend on the derivatives of  $\psi$. For this reason
$DD$ is not a derivation but an algebraic 2-form called the curvature 2-form $F$ \BE
 D D \psi = F \;\psi \;\;\hbox{with}\;\;\;
 F = d A + A A\,.
\EE If we develop all the indices in the $ A A$ term we find \BE
 A A = dx^{\mu} dx^{\nu} \;\; A_{\mu}^{a} A_{\nu}^{b} \;\; \LX_a \LX_b\,.
\EE
Since the $dx^{\mu}$ anti-commute and the $A_{\mu}^a$ commute, this expression is anti-symmetric
in the $(a,b)$ indices and therefore only involves the commutator of the $\LX _a$ matrices. Using the closure of the Lie algebra, we can rewrite $A A$ as a linear combination of the $\LX _a$ matrices, and see that $F$ is also valued in the adjoint representation of the Lie algebra $\ALG_0$.
\BE
 F = d A + A A = (d A^c + \demi A^a A^b f_{a b}^c) \LX _c = F^c  \LX _c\,.
\EE
The definition $DD = F$ only makes sense if the product of 3 or more $D$ is associative:
\BE
 D D D \psi = (D D) D  \psi = D (D D)  \psi \iff F D \psi  = D ( F \psi )\,.
\EE
These equations are equivalent to the Bianchi identity:
\BE
 d F + [ A , F ] = 0\,.
\EE
Expanding out all the indices, the terms in $A\;dA$ cancel because the 2-form $d A^a$ commutes
with the 1-form $A^b$. However the term trilinear in $A$
\BE
 dx^{\mu} dx^{\nu} dx^{\rho} \;\;A_{\mu}^{a} A_{\nu}^{b} A_{\rho}^{c} \;\; [\LX_a,\;[ \LX_b,\; \LX_c]]
\EE
vanishes if and only if the $\LX$ matrices satisfy the Jacobi identity, that is, if and only if
the connection $A$ is valued in the adjoint representation of a Lie algebra.

\paragraph{Theorem 1}
The Yang-Mills covariant differential $D = d + A$ is associative if and only if the connection
 1-form $A$ is valued in the adjoint representation of a Lie algebra.
\BE
 D\;\hbox{is associative} \iff \hbox{Bianchi} \iff \hbox{Jacobi}.
\EE
This theorem is implicit in the original paper of Yang and Mills \cite{YM54}, and
possibly even in the much earlier work of Elie Cartan, but I do not
know where it was first stated explicitly.

\section{The BRS differential}

The material in this section is also known, but will be needed in section 6.
The BRS differential $s$ controls the renormalization process of the
standard Yang-Mills theory \cite{BRS75}. The most economical way to introduce the
BRS operator and the Faddeev Popov \cite{FP67} ghost field $c$ is to
extend  the differential $d$ (2.10) and the Yang-Mills connection 1-form A (2.11)
as
 \BE
\dHS = d + s \,,\qquad \AHS = A + c\,.
 \EE 
We assume that the $c$ ghost anti-commutes with all 1-forms and we construct 
once more as in (2.12, 2.14) the covariant differential $\DHS$ and the curvature 2-form $\FHS$:
\BE 
\DHS = \dHS + \AHS  \,,\qquad \FHS = \dHS \AHS + \AHS \AHS\,. 
\EE
We now postulate the Maurer-Cartan horizontality condition:
\BE 
\FHS = F\,.
\EE
This equation must be understood as a constraint which defines
the action of $s$ on $A$ and $c$. Substituting (2.14) and (3.2) in (3.3),
we find:
\BE
s A = - d c - [A , c] \,,\qquad
s c = - \demi [c , c] \,.
\EE 
If we write explicitly the $\LX$ Lie algebra
matrices as in (2.11), we see that (3.4) includes only the commutator
of the $\LX$ matrices (2.7) and can be written as 
\BE s A^a = - d c^a -
f^a_{bc} A^b c^c \,,\qquad
s c^a = - \demi f^a_{bc} c^b c^c \,.
\EE 
We define the action of $s$ on higher forms by the
condition $s d + d s = 0$. We may then verify that the Jacobi
identity implies the consistency condition $s^2 = 0$.

\paragraph{Theorem 2}
The Yang-Mills-Faddeev-Popov covariant differential $\DHS = \dHS + \AHS$ is associative
if and only if the connection
 1-form $\AHS$ is valued in the adjoint representation of a Lie algebra, and if so
the BRS operator $s$ is nilpotent.

\BE
  \hbox{Bianchi} \iff \hbox{Jacobi} \iff s^2 = 0 \,.
\EE

 This theorem is implicit
in the work of Becchi, Rouet and Stora \cite{BRS75}, who only recognized the
variation $sc$ as the parallel transport on a Lie group. To the best
of my knowledge, the complete identification of the 2 BRS equations
with the horizontality conditions of Cartan was first proposed in my
own work \cite{TM80}.

\section{The Connes-Lott 2-points non commutative differential algebra}

This section recalls our last building block.
In their seminal paper \cite{CL90,CL91}, Connes and Lott introduced the non
commutative differential geometry of a discrete space with just 2
points. They consider 2 complementary projectors:
\BE 
\EX + \EXB = 1\,,\qquad
\EX \EXB = \EXB \EX = 0 \,,\qquad 
\EX \EX =\EX  \,,\qquad  
\EXB \EXB =\EXB \,, 
\EE 
and define their formal
differentials $\delta \EX$ and $\delta \EXB$. The interesting
observation is that the Leibnitz rule, together with the properties
of the projectors, imply non commutativity of the resulting
differential algebra: 
\BE
 \delta (\EX + \EXB) = \delta (1) = 0 =>
\delta \EXB = - \delta \EX \,,
\EE
 
\BE
 (\EX + \EXB) \;\delta \EX =
\delta \EX = \delta (\EX^2) = \delta \EX \;\EX + \EX \;\delta \EX =>
\EXB \;\delta \EX = \delta \EX \;\EX  \,,
\EE

\BE
 (\EX + \EXB) \;\delta \EXB =
\delta \EXB = \delta (\EXB^2) = \delta \EXB \;\EXB + \EXB \;\delta \EXB =>
\EX \;\delta \EX = \delta \EX \;\EXB  \,,
\EE
i.e., commuting with $\delta \EX$ exchanges the 2 projectors.
In this formalism, a function $f$ is a vector in the complex algebra with generators
$\EX$ and $\EXB$:
\BE
  f = g\; \EX + h \;\EXB \,,
\EE
 a connection 1-form $\omega$ is an object linear in $\delta \EX$:
\BE
 \omega = u\;\EX \;\delta \EX + v \; \EXB \; \delta \EX \;,
\EE
and the corresponding curvature $F$ can be written as: 
\BE
 F = \delta\omega + \omega \omega = (u - v + u v) \; \delta\EX\;\delta\EX \,.
\EE
 The curvature vanishes for the Cartan connection:
\BE
 \omega_0 = 2\;\EX\;\delta\EX - 2 \;\EXB\;\delta\EX \,.
\EE
Up to here, we exactly followed \cite{CL91}, but from now on our treatment
is original. These equations look more familiar if we
introduce the chirality operator $\chi$: 
\BE
 \chi = \EX - \EXB \;\;=>\;\;\chi^2 = 1\,,\qquad \chi\EX = \EX \chi\,, \qquad \chi\;\delta \chi = - \delta \chi \;\chi\,.
\EE
 Now $1$ and $\chi$ are the standard generators of the 2 elements
multiplicative group and the same flat connection now takes the
standard form of the Cartan left invariant 1-form:
\BE
 \omega_0 =
\chi\;\delta \chi = \chi^{-1}\;\delta \chi \,,
\EE
\BE
F_0 = \delta
\chi \delta \chi + \chi \; \delta \chi\;\chi \delta \chi = \delta
\chi \delta \chi - \delta \chi \; \chi \; \chi \; \delta \chi = 0 \,.
\EE
Forms and functions do not commute and we can define the covariant differential $\Delta$:
\BE
\Delta \; f = \delta\; f + \demi (\omega_0 \; f - f \; \omega_0)\,,
\EE
\BE
\Delta \; \omega = \delta\; \omega + \demi (\omega_0 \; \omega + \omega \; \omega_0)\,,
\EE
and so on, alternating the signs for higher forms.
Expanding all terms, and using (4.5, 4.6, 4.9), we verify that, for any function $f$, and
any p-form $\omega$, we have:
\BE
\Delta \; f = \Delta \; \omega = 0 \,. 
\EE
In other words, the differential $\delta$ is algebraic.

\section{The new Chiral-Yang-Mills connection}

Consider again a set of left and right spinors
$(\psi_{L}^{\alpha} , \psi_{R}^{\overline{\alpha}})$
as defined in (2.1).
Consider the 2 by 2 matrix notation corresponding to the left-right
splitting of the spinors. The top-left diagonal block is really of size $m m$, 
the bottom-right of size $ n n$. We define the left and right projectors
$\EX$ and $\EXB$ and the chirality operator $\chi$:
\begin{eqnarray}
\EX :  \psi_L \rightarrow \psi_L \,,\qquad \EXB  :  \psi_R  \rightarrow \psi_R \,,\qquad \qquad
\nonumber\\
\nonumber\\
\EX = \pmatrix { 1 & 0 \cr 0 & 0}\,,\qquad
\EXB = \pmatrix { 0 & 0 \cr 0 & 1}\,,\qquad
\chi = \EX - \EXB =  \pmatrix { 1 & 0 \cr 0 & -1}\,,
\end{eqnarray}
and the matrix 1-forms $\GX$ and $\GXB$:
\BE
\GX = \pmatrix { dx^{\mu}\;\sigma_{\mu} & 0 \cr 0 & dx^{\mu}\;\overline{\sigma}_{\mu}}\,,\qquad
\GXB =  \pmatrix { dx^{\mu}\;\overline{\sigma}_{\mu} & 0 \cr  0 & dx^{\mu}\;\sigma_{\mu}}\,.
\EE
Using (2.5,2.9) we have:
\BE
\GX \GXB = dx^{\mu} dx^{\nu} ( J_{\mu \nu}^L \;\EX + J_{\mu \nu}^R \;\EXB)\,.
\EE
The main point of this
paper is to generalize the Yang-Mills connection 1-form of equation (2.11) into the
chiral-Yang-Mills connection 1-form $\AHH$:
\BE
 \AHH = A + \Gamma \Phi\,.
\EE 
where the new term $\Gamma \Phi$ connects the left and the right Fermions.
Let the $a,b,c..$ alphabet labels the charges of $A$, and the
$i,j,k...$ alphabet labels the charges of $\Phi$:
\BE
A = A^a\;\LX_a\,,\qquad\Phi = \Phi^i\;\LX_i\,. 
\EE
In the 2 by 2 matrix notation (5.1), the $\LX_a$ and $\LX_i$ matrices are 
respectively block diagonal and anti-diagonal:
\BE
\Lambda_a = \pmatrix { \MX_a & 0 \cr 0 & \MXB_a}\,,\qquad
\Lambda_i =  \pmatrix { 0 & \MX_i  \cr \MXB_i & 0 } \,,
\EE
and we have
\BE
\GX \; \Lambda_i =  \Lambda_i \;\GXB \,,\qquad \GXB \; \Lambda_i =  \Lambda_i \;\GX \,.
\EE
Let us now study the  spin structure of  $A$ and $\Phi$, i.e. the way the Lorentz 
index $\mu$ appears in these definitions. $A$ is a 1-form linear in $dx^{\mu}$. 
Explicitly:
\BE
 A = A^a\;\LX_a = dx^{\mu} \; A_{\mu}^{a}(x) \;(\MX_{a \beta}^\alpha \;\;\EX + \overline{\MX}_{a  \overline{\beta}}^{\overline{\alpha}} \;\;\EXB)\,.
\EE
Therefore the function $A_{\mu}^a(x)$ is as usual a vector field. However, we have
to remember that the only way to connect a left to a right spinor of the $SO(1,3)$ Lorentz
algebra is to use the Pauli matrices (2.2-2.4) which, by definition, carry a $_{\mu}$ index.
As a result, the Pauli matrices $\sigma_{\mu}$ steal the $_{\mu}$ index usually carried by
the gauge field and the $^i$ charge is now carried by a field without Lorentz index, i.e.
by a scalar field $\Phi^i(x)$:
\BE
\GX \; \Phi =  \GX \; \Phi^i \; \LX_i  =dx^{\mu}\;\Phi^i(x)\;(\MX_{i  \overline{\beta}}^\alpha \;\;\sigma_{\mu} +
\overline{\lambda}_{i \beta}^ {\overline{\alpha}} \;\;\overline{\sigma}_{\mu})\,.
\EE
Consider now the $\MX$ and $\overline{\lambda}$ matrices which act on the 
charges $\alpha, \beta$ or
$\overline{\alpha}, \overline{\beta}$ of the spinors. In Yang-Mills theory, the $\LX$
matrices were chosen to represent a Lie algebra. In our
generalization, we make no a priori assumptions on the $\MX$
matrices, we will deduce their structure from the consistency
conditions on the connection 1-form $\AHH$. At this point,
$(\MX_a,\overline{\MX}_a,\MX_i,\overline{\MX}_i)$ are just defined
as complex valued matrices of respective dimensions $(mm,nn,mn,nm)$
where $m$ and $n$ are the numbers of left and right spinors. This
was the approach of Gell-Mann when he was studying the strong
interactions. He wanted to generalize the 3  $\lX$ matrices of the $SU(2)$
(proton,neutron) isospin of Wigner and include the strange particles. He successively
tried to use 4,5,6, or 7 matrices, before hitting the magic
eight-fold way. Independently, Ne'eman, who had read Dynkin, started
from the $G_2$ star of David, and then restricted it to $A_2 =
SU(3)$ flavor \cite{N61}.

Notice that by itself  $\GX$ (5.2) or $\Lambda_i$ (5.6) cannot act on the $\psi$, because
$\GX$ changes the chirality of the Fermions without changing the $\alpha$ charges 
and $\Lambda_i$ 
changes the $\alpha$ charges into $\overline{\alpha}$ charges without changing the chirality. 
Only the product $\GX \Lambda_i$ behaves correctly.
We now construct as in (2.12) the chiral-Yang-Mills covariant differential
\BE
\DHH = d + \AHH = d + A + \GX \Phi
\EE
and study as in (2.14) the repeated action of $\DHH$. Again $\DHH \DHH$ is a 2-form:
\BE
 \DHH \DHH \;\; \psi = \FHH \;\;\psi\,,
\EE
\BE
\FHH = d \AHH + \AHH \AHH = d A + A A + (A \;\GX \Phi + \GX \Phi \;A) + \GX \Phi \;\GX \Phi\,.
\EE
If we develop (5.12) in its (5.5) components, we find successively that
\BE
  A A = dx^{\mu} dx^{\nu} \;\; A_{\mu}^{a} A_{\nu}^{b}  \;\;\Lambda_a \Lambda_b
\EE
is anti-symmetric in $(a,b)$ (2.15), so that the $\Lambda_a$ matrices must be closed under commutation
\BE
  [\LX_a , \LX_b ] = \LX_a  \LX_b - \LX_b  \LX_a = f_{a b}^c \; \LX_c \, .
\EE
Then, in $(A \;\GX \Phi + \GX \Phi \;A)$, 
$A$ and $\GX$ are both 1-forms, and anti-commute. 
This produces a commutator which must close on the $\Lambda_i$:
\BE
 A  \;\;\GX \Phi + \GX \Phi  \;\;A = - \GX A^a \Phi^i  (\LX_a  \LX_i - \LX_i  \LX_a) =>  [\Lambda_a , \Lambda_i] = f_{a i}^j  \;\;\Lambda_j \,.
\EE
Finally, we consider the term $\GX \Phi \;\GX \Phi$. Using (5.7) we find
\BE
\GX \Phi  \;\;\GX \Phi = \GX \GXB  \;\;\Phi^i \Phi^j  \;\;\Lambda_i \Lambda_j\,.
\EE
Since the scalar fields
$\Phi^i$ and $\Phi^j$ fields commute, this term is symmetric in $(i,j)$, 
hence the $\Lambda_i$ matrices must close under
anti-commutation:
\BE
 \{\LX_i , \LX_j \} = \Lambda_i \Lambda_j + \Lambda_j \Lambda_i = d^a_{i j}   \;\;\Lambda_a\,,
\EE
i.e. the $d^a_{i j}$ structure constants are symmetric in $i,j$.
Again, we require as in (2.17) the associativity of the triple covariant differential
\BE
 \DHH \DHH \DHH  \;\;\psi = (\DHH \DHH) \DHH   \;\;\psi = \DHH (\DHH \DHH)   \;\;\psi \iff  \FHH \DHH \psi = \DHH (\FHH  \psi )  \,.
\EE
These equations are again equivalent as in (2.18) to the Bianchi identity
\BE
 d \FHH + [ \AHH , \FHH ] = 0\,.
\EE 
But this time, when we expand out the term trilinear in $\AHH$,
we find that the $\Lambda$ matrices must satisfy the super Jacobi
identity: 
\BE
 [ \LX_a , [\LX_b , \LX_c ] ] +  [ \LX_c , [\LX_a , \LX_b ] ] + [ \LX_b , [\LX_c , \LX_a ] ] = 0 \,,
 \nonumber \EE \BE
 [ \LX_a , [\LX_b , \LX_i ] ] +  [ \LX_i , [\LX_a , \LX_b ] ] + [ \LX_b , [\LX_i , \LX_a ] ] = 0 \,,
 \nonumber \EE \BE
 [ \LX_a , \{\LX_i , \LX_j \}] -  \{ \LX_j , [\LX_a , \LX_i ]\} + \{ \LX_i , [\LX_j , \LX_a ]\} = 0 \,,
 \nonumber \EE \BE
 [ \LX_i , \{\LX_j , \LX_k \}] +  [ \LX_k , \{\LX_i , \LX_j \}] + [ \LX_j , \{\LX_k , \LX_i \}] = 0 \,.
\EE
In other words, the $\Lambda_a$ and $\Lambda_i$ matrices form a Lie-Kac
super-algebra, where the $a$ charges label the even generators of the Lie sub-algebra 
(5.14 and 5.20 are identical to 2.7 and 2.8)
and the $i$ charges label the odd generators. 

\paragraph{Theorem 3}
The chiral-Yang-Mills covariant differential $\DHH = d + \AHH$ is associative if and only if 
the Fermions are graded by their chirality and the connection
 1-form $\AHH$ is valued in the adjoint representation of a Lie-Kac super-algebra.

\BE
 \DHH\; \hbox{is associative} \iff \hbox{Bianchi} \iff \hbox{super-Jacobi}.
\EE
As far as we know, this theorem is new.

\section{Non-commutative differential geometry of the chiral Fermions}

 In the chiral Fermion space, we can represent the Connes-Lott
2-point non commutative differential algebra as follows. The chirality operator (4.9) of the
2-point geometry is represented by the chirality operator (5.1) of the Fermion
space. The $\delta$ differential (4.2) is represented by the super-commutator with 
$\GX$ (5.2) times a fixed odd generator $v$ of the super-algebra (5.24)
\BE
 \delta = - [ \;\GX v, ] \,,\qquad \delta \psi = - \GX v \;\psi \,.
\EE
If we directly compute, in the 2 by 2 matrix notation (5.1,5.2) , the 
Cartan 1-form $\omega_0$ (4.10), using our new definition of the $\delta$ variation 
of $\chi$, we find
\BE
\omega_0 = \chi \; \delta \chi = 2  \;\GX v \,.
\EE
We recover as observed in (4.14) that $\Delta = \delta + \demi [\omega_0, ]$ vanishes on
$v$, $\EX$, $\chi$; $\omega_0$ has vanishing $\delta$ curvature (4.11); and $\delta^2$ 
vanishes on the chirality operators $\chi$ and $\EX$
\BE
 \delta \; \omega_0 = - \omega_0 \;\omega_0 \iff
  \delta \; \GX v  = - \{\GX v,\;\GX v\} \,,\qquad
  \delta^2\;\chi = \delta^2\;\EX = 0 \,.
\EE
Thanks to the super-Jacobi identity (5.23), $\delta^2$  also vanishes on $\GX v$ itself:
\BE 
	\delta^2 \;\GX v = - 2 \;[\GX v,\;\{\GX v,\; \GX v\}] = 0 \,.
\EE
These equations look very much like the BRS equations of 
section 3, except that the variation of $\GX v$ does not involve the usual 
factor $- 1/2$ found in (3.5), hence $\delta^2$ does not vanishes on $\psi$ and we have:
\BE
\delta^2 \; \psi = \demi \GX \GXB \{v,\;v\} \; \psi \,.
\EE

Let us now take advantage of the similarity between $s$ and $\delta$. We propose to mix the chiral (5.4) and the 
BRS (3.1) extensions of the Yang-Mills connection 1-form (2.11) and of the 
corresponding differentials and to
consider the new chiral-Yang-Mills-Higgs-Kibble-Faddeev-Popov-Connes-Lott connection 1-form $\AT$
\BE
\AT = \AHH + \cHH = \AHS + \GX \PhiHS 
\,,\qquad
\dT = \dHH + s = \dHS + \delta  
\,,
\EE
where we either decompose between gauge and ghost fields 
\BE
\;\;\; \AHH = A + \GX \Phi  \,,\qquad \cHH = c + \GX v \,,\qquad \dHH = d + \delta \,,
\EE 
or between the even and odd generators of the super-algebra
\BE
\AHS = A + c  \,,\qquad \PhiHS = \Phi + v \,,\qquad \dHS = d + s \,.
\EE
We now define the covariant differential $\DT$ and the curvature $\FT$ by
the usual formulas
\BE
\DT = \dT + \AT 
\,,\qquad
\FT = \dT \AT + \AT \AT \,.
\EE
However, if we compute $\DT^2$ we find a term $\delta^2$ which does not vanish (6.5).
Hence, in contradistinction to (2.14), we now have $\DT^2 = \FT + \delta^2$.
The Cartan horizontality condition (3.3) is really a consistency condition on $D^2$,
rather than on $F$, so the correct generalization is:
\BE 
\DT^2 = \FT + \delta^2  = \dT \AT + \AT \AT + \delta^2 = d\AHH + \AHH \AHH = \FHH = \DHH^2 \,.
\EE
The Cartan condition usually implies the exact gauge symmetry of the theory,
so we will not be surprised later to find that this $\delta^2$ correction 
controls the symmetry breaking of the standard model.
This equation is again a condition which defines the action of $s$ and $\delta$
on the gauge and ghosts fields. If we develop according to the different
gradations,  we find the usual definition (3.4) of $s$
\BE 
s A = - d c - [A , c] \,,\qquad
s c = - \demi [c , c] \,,\qquad
s \Phi = - [c , \Phi] \,,
\EE 
a novel definition of the action of $\delta$ on the gauge fields
\BE
\delta A = - \{\GX v , A\}  = - \GX [v , A] \,,\qquad
\delta \GX \Phi = - \{\GX v , \GX \Phi\} = - \GX \GXB \; \{ v , \Phi\} \,,
\EE
and the $s$ $\delta$ compatibility conditions
 \BE 
\delta c = - \{ \GX v , c\}  = - \GX [ v , c]\,, \qquad d v = s v = 0 \,.
\EE 
Again, we define the action of $\delta$ on higher
forms by the equation $\delta d + d \delta = 0$. The action of
$\delta$ on all fields is now defined algebraically. 
We can verify that $s^2 = 0$ as usual. However, we do not have
$\delta^2 = 0$, but thanks to the super-Jacobi identity (5.20-5.23)
we can verify that $\delta^2 = \demi \{\GX v,\; \GX v\}$
on all fields, generalizing (6.5).

The definition of $\delta$ deduced from the horizontality condition (6.10)
extends to $A$,$c$ and $\Phi$ (6.12-6.13) the geometrical condition (4.14)
that $\Delta$ should vanish on all fields. In other words, we have
exactly found the desired condition which usually holds true
on a principal fiber bundle, the $\delta$ ``vertical'' geometry (6.1)
exactly matches the non commutative differential geometry (4.2) of the 2-point 
finite space.

We also learn from (6.10) that the relevant gauge curvature 2-form is not
exactly $\FHH$ but $\FT = \FHH - \delta^2$. As mentioned before, this
object is also a 2-form if and only if we have the super-Jacobi
identity. If we develop we find:
\BE
\FT = dA + \demi[A , A] + \demi \GX \GXB (\{\Phi , \Phi\} - \{ v , v \}) - \GX D \Phi \,.
\EE
$\FT$ vanishes in the ``vacuum'' $A = 0, \Phi = \pm v$. 
It is therefore natural to rewrite $\FT$
in terms of the shifted field $\PhiHS = \Phi + v$ (6.8) and obtain
\BE
\FT = dA + \demi [A,\;A] + \GX \;D (\PhiHS - v) + \demi \;\GX \GXB\;(\{\PhiHS - v,\;\PhiHS - v\} - \{v,\;v\}) \,.
\EE
Finally, using the definition (6.3,6.12) of $\delta$, we can rewrite this last equation as
\BE
  \FT = (d + \delta) (A + \GX \PhiHS) + (A + \GX \PhiHS) (A + \GX \PhiHS)
\EE
which exactly matches the definition of Connes and Lott \cite{CL91}. 
Without the need to introduce anything else besides a generalized connection
1-form, we have recovered their main result.
The inclusion of the discrete differential
$\delta$ in the definition of the horizontal differential 
yields the Higgs-Kibble-Brout-Englert 
spontaneous symmetry breaking pattern
of the standard model. 

When we consider the differential $\dHH = d + \delta$ (6.7,6.16), we actually replace
the Yang-Mills geometry, where one considers a principal fiber
bundle with Minkowski space as the base and the Lie group as the fiber, by
the chiral-Yang-Mill geometry where the base space consists of 2
copies of Minkowski space, one populated by left movers and one by
right movers, reminiscent of super-string theory.

In this presentation, we have  identified the Connes-Lott discrete $\delta$
with the super-BRS operator associated to the odd generators of the
super-algebra, and $\GX v$ with the corresponding super-ghost.
Usually, the ghost of a tensor field has one index less than the
gauge field, for example the ghost of the
gauge 3-form of super-gravity is a 2-form,
the ghost of the graviton is a vector,
the ghost of the Yang-Mills vector is a scalar. Therefore, in the present case, 
where the 'gauge field' is a
scalar $\Phi$, it is perhaps not so surprising to find that its ghost is a 
constant $v$ satisfying
$d v = s v = 0$ (6.13).

Another striking fact is the interpretation of $v$ as the vacuum expectation
value of the $\Phi$ field. It may seem unusual to interpret the ghost as the vacuum
value of the gauge field. But in fact, in the conventional Yang-Mills theory, the
Yang-Mills-Faddeev-Popov connection $\AHS = A + c$ (3.1) defined on the principal bundle 
\cite{TM80b} never
vanishes, since its pull back on the gauge group is the left-invariant form. In the 
vacuum, we really have $A = 0$ but $\AHS = c$, i.e. the vacuum expectation
value of the full Yang-Mills connection is the Faddeev-Popov ghost form. Therefore,
the super-case $<\Phi> = v$ is the natural generalization of $<\AHS> = c$.

 To summarize, we have verified that:

\paragraph{Theorem 4}
The new chiral covariant differential $\DT = \dT + \AT$ is
associative if and only if the Fermions are 
graded by their chirality and the connection
 1-form $\AT$ is valued in the adjoint representation of a Lie-Kac super-algebra.
If so, the BRS operator $s$ is nilpotent and the square of the super-BRS Connes-Lott 
discrete differential $\delta$ is a constant 2-form. 
\BE
 \begin{array}{c} 
	 \hbox{Bianchi} \iff  \hbox{super-Jacobi} \iff 
\\
s^2 = s \delta + \delta s = 0 \;\;;\;\;\; \delta^2 = \demi\;\GX \GXB\;\{v,\;v\} \;.
\end{array} 
\EE
This theorem is new.

\section{Comparison with the Connes-Lott model}

The integration of the 2-point non-commutative differential
geometry in the chiral-Yang-Mills theory discussed
in the last section is very pleasing. It corresponds in
essence to the physical interpretation that Connes
and Lott give of their own work. Space time really has
2 chiral sheets, populated by left and right Fermions,
and the role of the Higgs field is to connect them, and 
give them a mass. Superficially, the Higgs potentials
computed here and in their paper agree. We both recover
directly the spontaneous symmetry breakdown of the
theory and subtract the constant term, so the
energy of the vacuum vanishes.
However, if we compare in
detail the Connes-Lott construction to the present one, things
become very different.

Whereas, in our new formalism, we found that a Lie
super-algebra, for example $SU(2/1)$, is necessary to
ensure the associativity of the covariant differential,
Connes and Lott do not consider this underlying
structure. Rather, they start from the universal
differential envelope of an associative algebra and
arrive at the Lie algebra $U(2).U(1)$ which acts separately 
on the 2 chiral subspaces. 
They do not really give a structure to the
$\Lambda_i$ odd matrices. The closest they come to our definition is
when they consider their junk ideal $J$. $J$ is needed to quotient out
the symmetric product of the $d x^{\mu}$, which exists in their
abstract universal differential envelope, but must be removed to
identify their $\pi(d x^{\mu})$ with a standard exterior form. But in the
interesting case describing the leptons, they find that the junk ideal
only contains in the 2-form sector the terms bilinear in two
Dirac matrices. The effect of quotienting by the Junk ideal 
therefore removes the trace of the term $\GX \GXB v v$ 
which appears in the curvature and lead to their formula
for the Higgs potential $V(\phi) = Tr (v^2 - Tr(v^2).Id)^2 ((\phi -
1)^2 -1)^2$. But since the term in $v^2$ occurs in the definition
of the curvature, we expect it to be valued in the adjoint
representation of the Lie algebra of the model.
However, if we look at the quantum numbers of the leptons,
we observe that the super-trace of the photon matrix vanishes, not
its trace, and Weinberg has explained in his original paper 
\cite{Wei67} that if we had in the algebra a second U(1) operator
besides the photon, then it would be coupled to the conserved lepton
number and be massless. But experimentally, this field does not exists. 
Therefore, by the argument of Weinberg, a traceless $v^2$ term cannot
be part of the curvature and in fact, a close reading of their Cargese 
lectures \cite{CL91} shows that Connes and Lott are aware of this difficulty.
We do not exactly know how to resolve this contradiction, but it may have
to do with the Wick rotation. Connes and Lott work in Euclidean space where
left and right Fermions are not well separated. In the end, they must rotate back
to Minkowski space. But if we think about it, the trace and the super-trace 
only differ by a relative sign when we trace succesively over the left and the
right spinors. May be the Wick rotation should involve in their formalism
a transmutation of the trace in to a super-trace. Notice also that in the Connes
Lott construction, a rather difficult calculation is needed to find,
at each degree $n$, the exact content of their junk ideal $J^{[n]}$. The
construction is needed to insure that they can work on the physical fields, 
i.e. on the equivalence classes obtained after division by $J$. 
On the other hand, in our formalism, the compatibility is insured 
to all exterior degrees by the Bianchi identity. 

On the technical side, we have completely by-passed their construction of 
the universal enveloping algebra, and the need to quotient by
the junk ideal or to invoke the Dixmier trace. We remain strictly
within the framework of standard Quantum Field Theory, gauge fields
and anti-commuting connection 1-forms. In this respect, our formalism is
much more economical. Although we adopt their 2-point non-commutative
differential geometry, the way we incorporate it in the new chiral-Yang-Mills
theory is different from their construction and leads to
a very different type of constraints on the acceptable gauge groups
and particle multiplets. In our construction, we must gauge
a finite dimensional classical Lie super-algebra, i.e. a
direct sum of members of the $SU(m/n)$, $OSp(m/n)$, 
$D(2/1;\alpha)$, $F(4)$ and $G(3)$ series of Kac. The Fermions
 must fall into finite dimensional representations of the 
chosen super-algebra, and be graded by chirality. The 
classification of the acceptable Connes-Lott models is completely
different. In a nutshell, Connes and Lott had the correct
physical intuition, but the wrong mathematical background !

To summarize, we found a way to import the 
fundamental concept of non-commutative differential geometry: the non 
commutative 2-point space which exchanges the 2 chiral sheets, and preserve
the main result of Connes-Lott: a geometrical interpretation of the 
Higgs field yielding automatically the symmetry breaking Higgs potential.
We did this, dropping their unusual formalism of the $K$-cycle, but
finding as a consistency condition, the existence of an underlying 
Lie-Kac super-algebra which grades the Fermions by their chirality.
Hence, as a cherry on the cake, we deduce
that the chiral Fermion geometry implies the
$SU(2/1)$ internal super-symmetry of Ne'eman and Fairlie 
\cite{N79,F79,NSF05}. 

\section{The topological term and the super-Killing metric}

As shown in section 5, the chiral-Yang-Mills  curvature 2-form $\FHH$
is valued in the adjoint representation of a Lie-Kac super-algebra:
\BE \FHH = \FHH^a\;\LX_a + \FHH^i\;\LX_i \,,
\EE
\BE
 \FHH^a = d A^a +
\demi f^a_{bc} A^b A^c + \demi \GX \GXB d^a_{ij} \Phi^i \Phi^j \,,
\EE
\BE
\FHH^i = \GX ( d \Phi^i + f^i_{aj} A^a \Phi ^j) \,.
\EE
 Notice that we
must use the equation $\LX_i \GX = \GXB \LX_i$ (5.7) before extracting the
$\LX_i$ matrices on the right. If we introduce the super-indices
$M,N$ which span the even an odd generators of the super-algebra 
\BE
\FHH = \FHH^M\;\LX_M \;\;;\;\;\; M = a, b, ..., i, j, ... 
\EE
 We can readily verify that the topological 4-form 
\BE
 \LAG_T =
\gHH_{MN}\;\FHH^M\;\FHH^N 
\EE
 is both closed, exact and $\delta$ exact:
\BE
 d \LAG_T =
\DHH \LAG_T = 0 \,, \qquad \delta \LAG_T = 0 \,,
\EE 
\BE
 \LAG_T = d (\AHH \FHH - 1/3 \AHH \AHH \AHH) \,,
\EE
 provided $\gHH_{MN}$ is the super-Killing metric of the super-algebra 
\BE
\gHH_{MN} = \demi \; STr (\LX_M\;\LX_N) = \demi \; Tr (\chi \; \LX_M\;\LX_N) \,.
\EE
These assertions depend on the super-Jacobi identity (5.20-5.23)
which implies that the quartic terms $\gHH_{MN}\;(\AHH \AHH)^M\;(\AHH
\AHH)^N$, hidden in the topological 4-form, actually vanish.
 Notice that the super-Killing metric $\gHH_{MN}$ (8.8)
is skew for the odd indices 
\BE
 \gHH_{ij} = - \gHH_{ji} \,. 
\EE
 This anti-symmetry exactly fits the odd part of (8.5)
\BE
\FHH^i \FHH^j = - \GX \GXB D \Phi^i \;D\Phi^j = - \FHH^j \FHH^i \,,
\EE
whereas the even part of (8.5) is symmetric as usual 
\BE
 \gHH_{ab} = \gHH_{ba} \,, \qquad \FHH^a \FHH^b = \FHH^b \FHH^a \,.
\EE
This concludes the study of the differential geometry of the chiral
Yang-Mills theory. We will analyze why the model is interesting
for particle physics in a separate paper.

\section{Discussion}

In Yang-Mills theory, the underlying symmetry group, which is truly
the product of the SO(3,1) Lorentz rotation group by the local gauge
group, say $SU(2) U(1)$, is factorizable. The 2 indices $a$ and
$\mu$ of the Yang-Mills field $A^a_{\mu}$ do not speak to each
other. Since the discovery by Ne'eman and Fairlie in 1979 of the
$SU(2/1)$ structure of the weak interactions, many authors tried to
integrate the super-algebra structure into Yang-Mills theory. The
stumbling block, which prevented this fusion, was the implicit
assumption that this new theory would also be factorizable. But
mixing left and right spinors at a given point of space time breaks
the Lorentz group. As we recalled in this paper, we can only mix
left and right in a given direction of propagation $x^{\mu}$ and
only using the $\sigma$ matrices.
 In the same way, internal super-symmetry between particle with the same
spin breaks the spin-statistics relation. For example, in 82 with
Yuval Ne'eman \cite{TMN82}, we tried to embed the $SU(2/1)$ $Z_2$ grading in
the $Z$ grading of exterior forms, anticipating on the Quillen
connection \cite{Q85}. But this did not work because the left spin 1/2 was
rotated into a left spin 3/2 state, and the non Abelian equations
were not integrable.

The factorization issue was addressed by Connes and Lott in 1990 \cite{CL90}.
However, it is very difficult to build upon their foundations, because they
seem to imply that the usual functional integral Lagrangian formalism of
Quantum Field Theory must be replaced by a completely new framework
based on a $K$-cycle and the Dixmier trace. Furthermore, their $\delta$ differential
fails to close unless they quotient by their junk ideal which has no
clear interpretation in QFT. Many papers followed which tried to
explain the Connes-Lott formalism to the physicists by throwing away
the Dirac matrices and reverting to the factorized framework. But,
as shown by \cite{PPS94}, they were throwing the baby with the bath and
can be ignored.

The new construction proposed in this paper is clenching the rotation group
and the internal super-symmetry in an indissoluble way. Neither the Dirac
one form $\GX$ (5.2), nor the odd matrix $\LX_i$ (5.6) can act alone on the Fermions. Only
their chiral product $\GX \LX_i$ is well defined and contribute to the connection
one form (5.4) as $\GX \LX \Phi$. We then found that the covariant differential (5.10)
is associative if and only if $A^a,\Phi^i$ is valued in the adjoint
representation of a super-algebra (5.24). Then we showed that the super commutator
with a fixed odd direction (6.1) implements the
2-point non-commutative differential 
differential of Connes-Lott (4.2) and leads as they found to spontaneous symmetry
breakdown (6.15), but without leaving the standard framework of Quantum
Field Theory. 
The super-Jacobi identity  (5.23) automatically implements
the consistency (6.17) of our definition (6.1) of the discrete differential $\delta$. 
This construction is much simpler and completely different from \cite{CL90},
and leads to a similar but not identical expression for
the Higgs potential and the choice of the $U(1)$ weak hyper-charge.
If this chiral geometry really exists at very high energy, may be we
should search for a compactification 
which would preserve down to 1000 Gev some of the left-right 
dialectic so characteristic of the 10-dimensional super-string.

Our new construction exactly fits the phenomenology
of the weak interactions. If we grade the Fermions by their chirality (5.24),
and consider the smallest simple super-algebra $SU(2/1)$,
we recover the phenomenological model of Ne'eman and Fairlie \cite{N79,F79,NSF05}.
This model appeared in 1979 as very promising. It requires
that the weak $U(1)$ hyper-charge should be supertracless,
automatically implementing the original choice of Weinberg \cite{Wei67}
who wanted to avoid a current coupled to the lepton number, and
explaining the non existence of any massless charged particle.
In addition. the irreducible representations of $SU(2/1)$ naturally describe
both the leptons \cite{N79,F79} and the quark \cite{DJ79,NT80}. However no setting
was found so far that could explain how to extend the Yang-Mills Lie
algebra into a super-algebra, and no relation was found between
chirality and the super-algebraic structure. We believe that the present paper
exactly fills these gaps and will come back in a
separate paper to the discussion of the $SU(2/1)$ model. 

\bigskip

\acknowledgments

It is a pleasure to thank John Lott for his gentle guiding through
non commutative geometry, Florian Scheck for an interesting
reprint, Pierre Fayet and Victor Kac, my dependable references 
in mathematics and physics, Gerard Menessier and Andre Neveu 
for many discussions and Yuval Ne'eman for
our long lasting collaboration on SU(2/1). We also
thank David Lipman for inviting us to his wonderful laboratory and
Danielle Thierry-Mieg for innumerable creative suggestions.

This research was supported in part by the Intramural 
Research Program of the NIH, National Library of Medicin.

\end{document}